# The avalanche instability of the orbital angular momentum in combined vortex beams

A. Volyar, M. Bretsko, Ya. Akimova, Yu. Egorov

*Physics and Technology Institute, V.I. Vernadsky Crimean Federal University,
Academician Vernadsky 4, 295007, Simferopol, Republic of Crimea, Russia.*
*volyar@singular-optics.org*

## I. Introduction

Among the diversity of current tasks of singular optics (see e.g. [1-4] and references therein), we focused attention on the problem of the propagation and transformations of combined singular beams. A scalar combined vortex beam consists of a set of partial singular beams carrying optical vortices with integer topological charges of different orders. Its state is conveniently characterized by the amplitudes and phases of the modebeams, which we will call the vortex spectrum. Knowing the vortex spectrum, it is easy to reproduce the orbital angular momentum (OAM) of the combined beam [5], which, as a rule, has a fractional value.

Analyzing the problem of combined beams, we conditionally divided it into three stages: the beam shaping, its propagation and transformations, and detection of the combined beam properties. It is most convenient to form combined beams using diffractive optical elements (DOE) based on digital holographic grating (see e.g. [6-9] and references therein). It was this approach that stimulated the widespread use of combined beams in systems of optical data transmission [10], in quantum computers, optical cryptography [11], and in micro-and-macro objects manipulations [12]. The structural evolution of combined beams with fractional OAM during their propagation depends on the vortex composition. For example, a beam shaped by an imperfect spiral phase plate splits immediately into a complex array of optical vortices [13-15]. At the same time, certain changes in the vortex composition at the initial plane can cause a complex vortex pattern to transform into a strictly ordered structure during propagation [16,17]. Structural instability of combined beams with fractional OAM can be eliminated in a space variant birefringent medium of photonic crystals. Such structurally stable combined beams are called optical quarks [18].

Significant attention is attracted to the measurement technique of the vortex composition. The main emphasis in this question is placed on the techniques of holographic gratings [8,19-21] and special refractive elements [22]. The basis of these techniques is the mode beams sorting in accordance with the integer OAM numerically equal to the vortex topological charge. The sorting of optical vortices is inevitably associated with the perturbation of the combined beam microstructure and information losses. An alternative approach is based on the intensity moments technique [23,24]. For the first time this technique was partially used in Ref.[25] to measure the OAM in singular beams with fractional topological charge. The authors of this technique determined the OAM measuring second-order intensity moments. Recently, the authors of Ref.[26] expanded this approach by measuring higher-order intensity moments. As an example, the authors demonstrate resonant bursts of the OAM in vicinity of the integer values of the vortex topological charges. The fact is that the authors of Ref. [16,27] predicted the presence of sharp OAM "resonances" from zero to large values of the OAM, which depend on the amplitude distribution of the vortex modes in a combined beam. However, it was experimentally shown in Ref.[21,25] that such resonant bursts are absent, both for small and for large values of the vortex topological charges. At the same time, to compare the results of these independent studies, it is first necessary to take into account the shape of holographic gratings that restore the beam and embed "resonant" properties in it. There have been no studies on this problem so that the question remains open. Thus, the aim of our paper is a theoretical and experimental study of

OAM resonances in the combined vortex beams restored by holographic gratings with weak perturbations of their structure.

The paper is organized as follows. In Section 2 we treat typical models of combined vortex beams with anomalous properties of the orbital angular momentum. The main attention is focused on a physical nature of the OAM resonances. Section 3 analyzes the properties of the complex amplitude decomposition coefficients in the basis of Bessel beams and Laguerre-Gauss beams that is responsible for the OAM avalanche instability. The experimental techniques for measuring the composition of vortex modes in the combined beams and the OAM transformations caused by perturbations of the holographic grating are discussed in Section 3.

## 2. The combined beam model

In the token paper [28], Berry showed that small perturbations of the spiral wave plate lead to a significant distortion of the beam structure due to the internal damage of the optical vortex. For example, a spiral wave plate of a $p-$ fractional step causes the birth of an optical vortex with a fractional topological charge $p$ that decays into infinite number of integer-order vortices. Following this idea, the authors of the papers [23,32] presented a new type of combined singular beams carrying fractional OAM. The approach was based on a model of conical plane waves in which the phase tracing around the cone axis is characterized by a fractional number $p$. The angular spectrum of the beam is written as

$$U(\phi, p) = \exp(i\, p\, \phi) = \frac{e^{i\pi p} \sin(\pi p)}{\pi} \sum_{m=-\infty}^{\infty} \frac{e^{im\phi}}{p-m}, \quad (1)$$

where $\phi$ is azimuthal angle in the wave vector space, $m = 0, \pm 1, \pm 2, \ldots$. In order to restrict the energy flow in the model and make the model convenient for experimental implementation, it was proposed [15–18] to put the conical beams of plane waves into a thin ring with a Gaussian distribution of amplitudes. Then the complex amplitude of the beam field in the initial plane takes the form

$$\Psi(R,\varphi) = 2 N G(R) e^{ip\pi} \sin p\pi \sum_{m=-\infty}^{\infty} \frac{i^m I_m(\Omega R)}{M_m(p-m)} e^{im\varphi}, \quad (2)$$

where $I_m(KR)$ is a modified Bessel function of the first kind, $R = r/w_0$, $N = \exp(-iK^2/2k z_0)$, $z_0 = kw_0^2/2$, $w_0$ is a beam waist radius, $G(R) = \exp(r^2/w_0^2)$. $r$ and $\varphi$ are the radial and azimuthal coordinates, $M_m$ is an amplitude coefficient, $\Omega$ stands for a scale beam parameter. Thus, the complex amplitude describes a combined singular beam in the basis of Bessel-Gauss beams (modes). The amplitude factor $\sin p\pi /(p-m)$ selects the modes in the combined beam. This can be seen from the relations

$$\lim_{p \to m} \frac{\sin \pi p}{\pi(p-m)} = \cos m\pi \quad \text{and} \quad \lim_{p \to \infty} \frac{\sin \pi p}{\pi(p-m)} = 0, \; p \neq m. \quad (3)$$

The effect of this factor is most clearly manifested in the example of the beam OAM. With that end we use the standard representation of the longitudinal component of the OAM per photon [28] and obtain

$$\ell_z = \frac{i\langle \Psi | \partial_\varphi | \Psi \rangle}{\langle \Psi | \Psi \rangle} = \sum_{m=-\infty}^{\infty} \frac{m I_m(\Omega^2)}{M_m(p-m)^2} \bigg/ \sum_{m=-\infty}^{\infty} \frac{I_m(\Omega^2)}{M_m(p-m)^2}, \quad (4)$$

Such an OAM representation depends on the amplitude coefficients $M_m$ in a beam superposition. For example, if the complex amplitude (2) is represented in the normalized basis of Bessel beams (with $M_m^2 = \langle \Psi | \Psi \rangle$) then the OAM is written in the form [21,30]

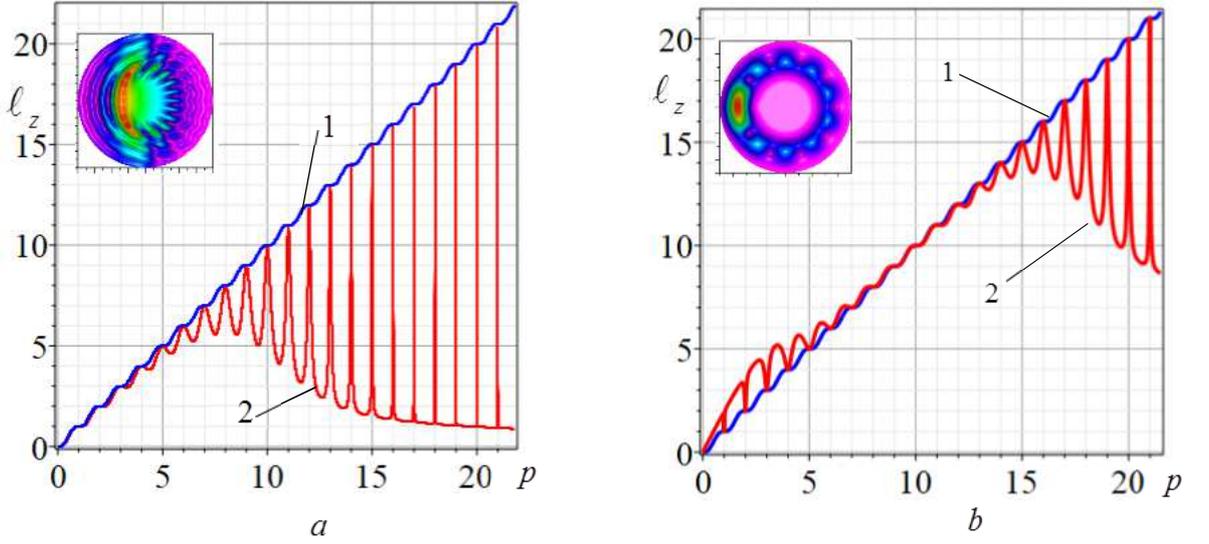

Fig.1 The dependence of the OAM $\ell_z$ on a perturbation parameter $p$ for the combined Bessel-Gaissian beams: a) curve 1 is calculated according to Eq. (5) while the curve 2 is the result of Eq.(4), curve 2 is plotted according to Ea. (5) $\Omega = 10$, ; (b) $\ell_z(p)$ for the Laguerre-Gaussian combined beams, Eq.(8), $\Omega = 10$. The callout shows the theoretical intensity distribution of the cross section of a disturbed beam at the waist plane with $p = 19.5$.

$$\ell_z = \sum_{m=-\infty}^{\infty} \frac{m}{(p-m)^2} / \sum_{m=-\infty}^{\infty} \frac{1}{(p-m)^2}. \tag{5}$$

The dependence of the OAM $\ell_z$ on the vortex parameter $p$ is illustrated by the smooth curve 1 in Fig. 1a. This theoretical background was experimentally confirmed for small values of the parameter by the authors of Ref. [14,27]. The alternative case for $M_m = 1$ represented by Eq. (4) is illustrated by curve 2 in Fig. 1a. At low values of the $p-$ parameter, curves 1 and 2 almost coincide. However, at large values $p$, resonance bursts are observed in the spectrum $\ell_z(p)$ [15]. The height of the resonances and the width of the resonant contour depend both on the parameter $p$ and the scale parameter $\Omega$.

We use here the terms "resonant bursts and dips" and "spectral" OAM curve $\ell_z(p)$ without quotes, based on the following reasons. The dependence of the OAM on the beam parameter is the result of the interference between vortex beams (modes) represented by the Fourier series. The shape of the curve $\ell_z(p)$ is given by the average values of the angular momentum operator $i\partial_\varphi$, such as $\ell_z(p) = i\langle\Psi|\partial_\varphi|\Psi\rangle$, which depends both on the eigenvalues spectrum $m$ of the operator and the squared amplitudes $C_m^2$ of the vortex beams. A monotone growth of the curve $\ell_z(p)$ can be broken by sharp bursts in the OAM, when the $p-$ parameter coincides with the eigenvalues of the operator, i.e $p = m$. These bursts and dips we call the OAM resonances.

Let us consider a new model of a combined beam endowed with OAM resonant properties. We write the angular spectrum of the combined beam in the form

$$U(k_\perp, \phi) = \frac{e^{ip\pi} \sin \pi p}{\pi} \sum_{m=-\infty}^{\infty} \frac{(\Omega r k_\perp)^m}{M_m m!} \frac{e^{im\phi}}{p-m} e^{-\frac{k_\perp^2 w_0^2}{4}}. \tag{6}$$

Then the complex amplitude of the combined beam is written as

$$\Psi_P(r,\varphi,z=0) = \frac{e^{ip\pi} w_0 \sin \pi p}{2\pi} \sum_{m=-\infty}^{\infty} \frac{(\Omega R)^{|m|}}{|m|!} \frac{e^{im\varphi}}{p-m} e^{-\frac{r^2}{w_0^2}}, \qquad (7)$$

provided that $M_m = \sqrt{\Gamma(m+1/2)/2^{(2|m|+3)/2}}$ The complex amplitude (7) is specified in the incomplete basis of Laguerre-Gauss beams with a radial index $n=0$ and is also characterized by two parameters $p$ and $\Omega$. The complex amplitude (7) can be represented in a closed form via special functions [37]

$$\Psi_p = (-1)^p e^{ip\varphi} \frac{(\Omega R)^p \sin(p\pi)}{2\pi} \left\{ -\gamma(-p, \Omega R e^{i\varphi}) + \right.$$
$$\left. + \frac{p+2}{(p+1)^2} e^{i\varphi} \left[ {}_1F_1(p+1; p+2; \Omega R e^{i\varphi}) \right] \right\} e^{-R^2}, \qquad (8)$$

where we used the series [29]

$$\sum_{m=0}^{\infty} \frac{x^m}{m!(m-p)} = (-x)^p \gamma(-p, x), \quad \sum_{m=1}^{\infty} \frac{x^m}{m!(m+p)} = \frac{x}{p+1} {}_2F_2(1, p+1; 2, p+2; x)$$

$\gamma(a,x) = \Gamma(a) - \Gamma(a,x)$ is an incomplete gamma-function, $\Gamma(a)$ is a gamma-function and $\Gamma(a,x)$ is an additional incomplete gamma-function, ${}_2F_2$ stands for a hypergeometric function. The properties of hypergeometric beams have already been considered for beams presented in a factorized form, i.e. in a form where the functions of the angular $\varphi$ and radial $r$ coordinates are separated [35,36]. Vortex-beams with fractional topological charges are usually written in a more complex non-factorized form [18]. In fact, any superposition of vortex beams can be represented in a non-factorized form.

The non-standard shape of the beam vortex structure in Eq. (7) raises the question, what is the physical interpretation of the fractional parameter $p$? This question entails inevitably the following question: how is the parameter $p$ related to the topological charge $S_p$ of the combined beam? By definition [28], the total topological charge $S_p$ is a signed sum of all the vortices threading a large loop including the beam axis. As previously noted, Berry showed that the topological charge of the vortex beam diffracted by a fractional-step spiral phase plate changes almost linearly with the nearest integer to the fractional step $p$ (see Eq. (22) in Ref.[28]) . Developing this idea, the authors of the paper [20] confirmed experimentally this theoretical prediction. However, the authors of the papers [33] considering more complex cases have theoretically and experimentally showed that the topological charge and the OAM per photon of the combined beam are distinct physical quantities, and their numerical equality can coincide approximately only in particular cases. In Appendix, we estimate the total topological charge $S_p$ of the combined beam (7) containing a finite number $2N$ of mode beams with $m \in (-N, N)$ and show that this is defined by the maximum value of the vortex topological charge $|N|$ in the combined beam: $S_p = N$ for $p > 0$ and $S_p = -N$ for $p < 0$. However, as soon as the fractional number $p$ comes close to an integer value $p \to M$, $M \in (-N, N)$, the total topological charge becomes equal to this integer value $S_p = M$. This paradox is resolved if we consider the parameter $p$ as a perturbation acting on a vortex beam with a topological vortex charge $M$. A weak perturbation destroys the ground state of the beam in such a way that the initial intensity is redistributed among the set of mode beams in the range of topological charges from $-N$ to $N$. The type of the intensity redistribution is determined by the coefficients of the decomposition of the complex amplitude over mode beams (e.g. in Eqs (2) and (7)) and can be experimentally studied by measuring the amplitudes of the mode beams and the OAM of the combined beam. The problem of selective destruction of higher-order vortex beams was already addressed in the framework of the standard models of a turbulent medium [40-43]. The averaged effect of phase inhomogeneities of the turbulent medium in these models blurs fine resonant processes of the OAM. In the following, we will try to focus on the resonant properties of the OAM caused by perturbations of the holographic grating responsible for shaping combined beams.

Thus, let us consider peculiar properties of the OAM. The detailed analysis presented in Ref. [16,18,34] showed that that a combined vortex-beam with wished positions of singularities and the OAM can be shaped via combining vortex beams with different topological charges. In particular, the OAM of the vortex-beam (7) in our model is

$$\ell_z = \sum_{m=-\infty}^{\infty} \frac{m\Omega^{2|m|}}{|m|!^2 (p-m)^2} \bigg/ \sum_{m=-\infty}^{\infty} \frac{\Omega^{2|m|}}{|m|!^2 (p-m)^2}. \qquad (9)$$

The spectral curve $\ell_z(p)$ is represented by the curve 2 in Fig.1b. At first, as the perturbation parameter $p$ increases, a small resonant dips of the OAM from $\ell_z \approx 2$ to $\ell_z = 1$ arises in the interval $(0.5, 1.5)$, whereas sharp OAM dip from $\ell_z \approx 4$ to $\ell_z = 2$ is observed in the interval $p \in (1.5, 2.5)$. Then the spectral curve monotonously increases in the interval $p \in (5, 13)$ where it almost coincides with the curve 1 plotted according to Eq. (5). A further increase of the parameter $p$ is accompanied by a sequence of resonant bursts. We will discuss the physical reasons of this effect in details later (see Section 3) analyzing the spectrum of squared amplitudes. To determine the extreme values of the angular momentum, we find $\ell_{ext} = \lim_{p \to M} \ell_z$, $M = \pm 1, \pm 2, \ldots,$ as follows

$$\ell_{ext} = \lim_{p \to M} \frac{\sum_{m=-\infty}^{\infty} \frac{m\Omega^{2|m|}}{|m|!^2 (p-m)^2}}{\sum_{m=-\infty}^{\infty} \frac{\Omega^{2|m|}}{|m|!^2 (p-m)^2}} = \lim_{p \to M} \frac{\frac{M\Omega^{2|M|}}{|M|!^2 (p-M)^2} + (p-M)^2 \sum_{\substack{\infty < m < \infty \\ m \neq M}} \frac{m\Omega^{2|m|}}{|m|!^2 (p-m)^2}}{\frac{\Omega^{2|M|}}{|M|!^2 (p-M)^2} + (p-M)^2 \sum_{\substack{\infty < m < \infty \\ m \neq M}} \frac{\Omega^{2|m|}}{|m|!^2 (p-m)^2}} = M \qquad (10)$$

The series in the numerator and denominator converge for all $\Omega \in \mathbb{C}$ if $|p-m| \geq 1$. Note that the maxima of the OAM $l_{max} = M$ are replaced by its minima $l_{min} = M$, if $p < 12.4$ for $\Omega = 10$. The OAM minima are located between two maxima $M-1$ and $M$ at the points $p = M \pm 1/2$ for $p > 12.4$ and $\Omega = 10$. The height of the resonant bursts $h_r = \ell_{max} - \ell_{min}$ (or the depth of the dips) is illustrated in Fig.2. As the topological charge $M$ increases, the burst height rapidly grows and tends to $h_r \to M$.

The shape of the resonance contour is transformed with changes in both the perturbation parameter $p$ and the scale parameter $\Omega$, as can be seen in Fig.1a,b the curves 2. It is convenient to characterize the contour of OAM resonances by half width $\Delta p$, i.e. the value of the fractional parameter $p$ at which the OAM decreases by 2 times relative to the maximum value for the integer value in Eq.(4). Then the equation for the

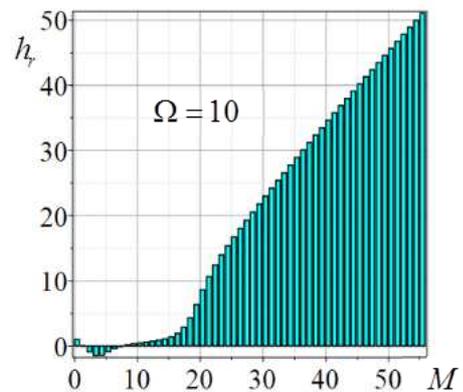

Fig.2 The resonant burst height $h_r(M)$, $\Omega = 10$

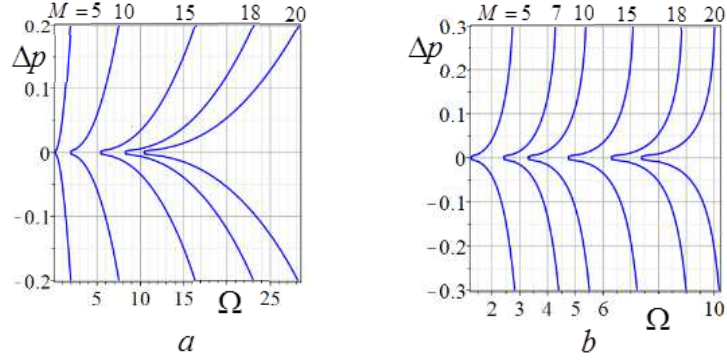

Fig.3 The dependence of the contour width $\Delta p$ on the scale parameter $\Omega$: (a) for the complex amplitude (3), for the complex amplitude (7).

half width of the resonances OAM (4) takes the form

$$\frac{1}{2}M = \sum_{m=-\infty}^{\infty} \frac{m I_m(\Omega)}{(M-\Delta p - m)^2} \Big/ \sum_{m=-\infty}^{\infty} \frac{I_m(\Omega)}{(M-\Delta p - m)^2} \qquad (11)$$

For the OAM (8), the equation for half width takes the form

$$\frac{1}{2}M = \sum_{m=-\infty}^{\infty} \frac{m \Omega^{2|m|}}{|m|!^2 (M-\Delta p - m)^2} \Big/ \sum_{m=-\infty}^{\infty} \frac{\Omega^{2|m|}}{|m|!^2 (M-\Delta p - m)^2}. \qquad (12)$$

The diagrams in Fig.3a,b show the families of curves $\Delta p(\Omega)$ for the integer values of topological charges $M$. Note that the front and back contour lines $\ell_z(p)$ are asymmetrical. It is property that is reflected in the upper and lower branches of the curve in Fig. 1a,b. On the upper horizontal line of Fig.3, there are integer values $M$ of the vortex topological charge, The width of the contour $\Delta p$ depends critically on the scale parameter $\Omega$ and the corresponding value of the maximum angular momentum $\ell_z = M$.

### 3. The vortex avalanche instability

The key part in shaping the OAM resonances is played by the choice of mode amplitudes in the combined beam (7):

$$C_m^2 = \frac{\Omega^{2|m|}}{|m|!^2 (p-m)^2} \Big/ \sum_{m=-\infty}^{\infty} \frac{\Omega^{2|m|}}{|m|!^2 (p-m)^2}. \qquad (13)$$

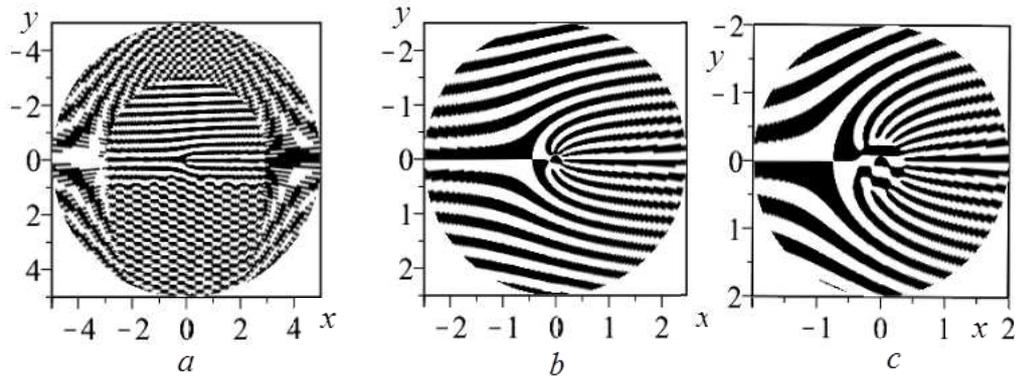

Fig.5 The holographic grating for different vortex beam states (a) $|2\rangle$, $\delta p = 0.01$, (b) $|10\rangle$, $\delta p = 0.001$, (c) $|15\rangle$, $\delta p = 0.001$; $\Omega = 10$

As a rule, a single optical vortex can be formed by a forked holographic grating. In turn, the mode amplitudes (13) can be embedded in the forked grating in the form of weak perturbations. Thus we assume that the binary holographic grating is matched with the beam field (7), so that any changes in the beam structure is associated with weak perturbations $\delta p$ of the grating relief. In this case, the perturbation $\delta p$ acting on the grating as the deviation $\delta p = M - p$ from the integer value $M$ of the nonperturbed vortex topological charge. We denote the beam initial state as $|M\rangle$. For example, a perturbed holographic grating of a standard combined beam is shown in Fig.4 for the perturbation $\delta p = 0.3$ of the state $|9\rangle$ while the OAM spectrum $\ell_z(p)$ of the combined beam is devoid of resonances (see Eq.(5) and the curve 1 in Fig.1a).

A completely different grating relief is associated with the mode amplitudes (13). The effect of the perturbation on the combined beam depends on both the initial vortex state $|M\rangle$ and the scale parameter $\Omega$. The binary grating relief is given by

$$T_P = signum\left[\cos\left(\arg\Psi_P - Qr\cos\varphi\right)\right], \quad (14)$$

where $Q$ is a scale parameter of the grating. For example, in order for the holographic grating to form a singular beam capable of changing appreciably its properties, we chose the scale parameter $\Omega = 7$, beam states $\langle 10|$ and $\langle 15|$. According to the diagram in Fig. 3b, the width of the OAM contour for the state $\langle 10|$ corresponds to overlapping neighboring contours while for the state $\langle 15|$, the half-width is approximately $\Delta p \approx 0.1$ that can serves for the experimental verification. Typical binary holographic grating is shown in Fig.5. The relief of the grating in Fig. 5b for $\delta p = 0.001$ in the state $|10\rangle$ is almost the same as that of the nonperturbed grating. Therefore, it can be expected that the beam OAM in the state $|10\rangle$ will be weakly subject to perturbations. However, the grating relief of the state $|15\rangle$ and $\delta p = 0.001$ in Fig.5c changes essentially. This means that one should expect the OAM of the initial state $|15\rangle$ breaks down under the action of a weak perturbation. At first glance, the perturbation $\delta p = 0.001$ of the state $|2\rangle$ should not cause significant structural changes of the grating. However, it turned out that even such a weak perturbation transforms radically the grating structure shown in Fig.6a. As a result, as can be seen from Fig.1c (curve 2) this weak perturbation causes a dip in the spectral curve.

To understand the physical causes of dips and bursts in the OAM spectrum $\ell_z(p)$, we turn to the intensity distribution $I_m \sim C_m^2$ over the beam modes. The dependences of the squared amplitudes $C_m^2$ on the topological charge $m$ for beam states $|10\rangle$, $|20\rangle$, $|50\rangle$ and $|200\rangle$ subjected to perturbations $\delta p$ are illustrated in Fig. 6-8.. In the initial state $|10\rangle$, the main beam intensity is concentrated in the single vortex mode with $C_m^2 = 1$. Perturbations in the interval $\delta p = 0.2 \div 0.5$ (Fig.6 a-d) does not cause significant intensity redistribution between modes in the interval where the OAM changes almost linearly with increasing the parameter $p$ (compare with Fig.1b).

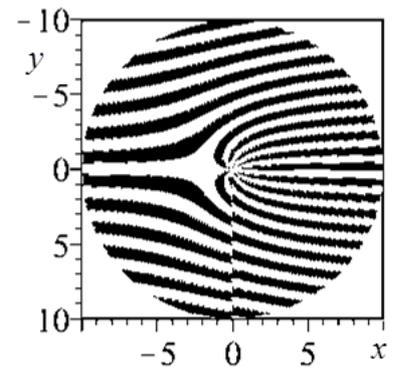

Fig.4 Perturbed holographic grating of the standard combined beam with the angular spectrum (1); $\delta p = 0.3$, $\langle M| = |9\rangle$

A completely different situation arises with vortex-beams whose topological charges are higher than $M > 15$ and $\Omega = 10$. A slight perturbation $\delta p = 0.02$ of the beam $|20\rangle$ causes a breaking-down of the vortex state so that the beam intensity is drastically redistributed among an avalanche of mode beams in a broad spectral region of states $\langle M |$ shown in Fig.7a. As the perturbation increases, the avalanche of states grows up. The modes are grouped in the vicinity of $|12\rangle$ and $|-8\rangle$ states. Most of the intensity is pumped into satellites around the state $|12\rangle$ (Fig.7b-d). The larger the fractional $p$-value, the more intensity is pumped to the side mode satellites with intensity peaks located at the region of positive and negative topological charges $m_{max}$.

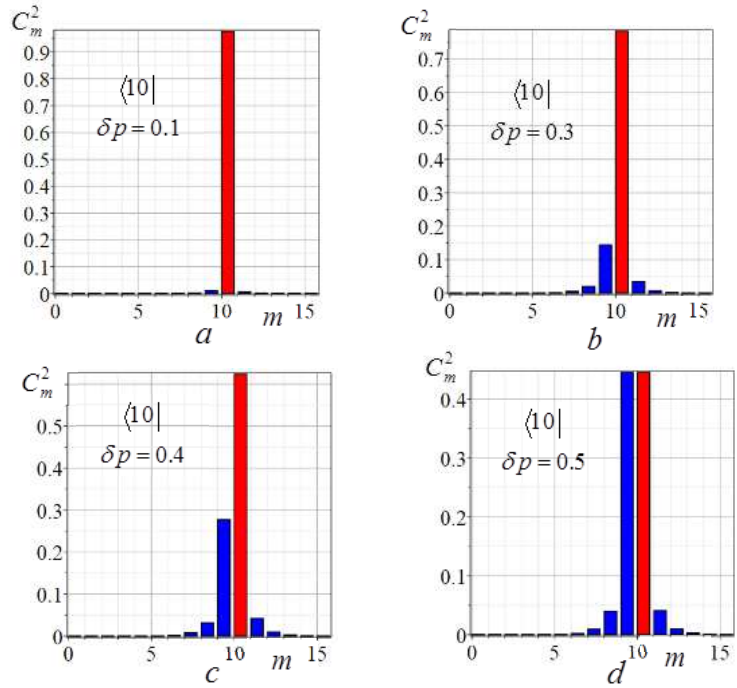

Fig. 6 The computer simulation of the squared amplitude $C_m^2$ of the vortex-beam (7) in the state $|10\rangle$ (red color) perturbed by: (a) $\delta p = 0.1$, (b) $\delta p = 0.3$, (c) $\delta p = 0.4$, (d) $\delta p = 0.5$; $\Omega = 10$

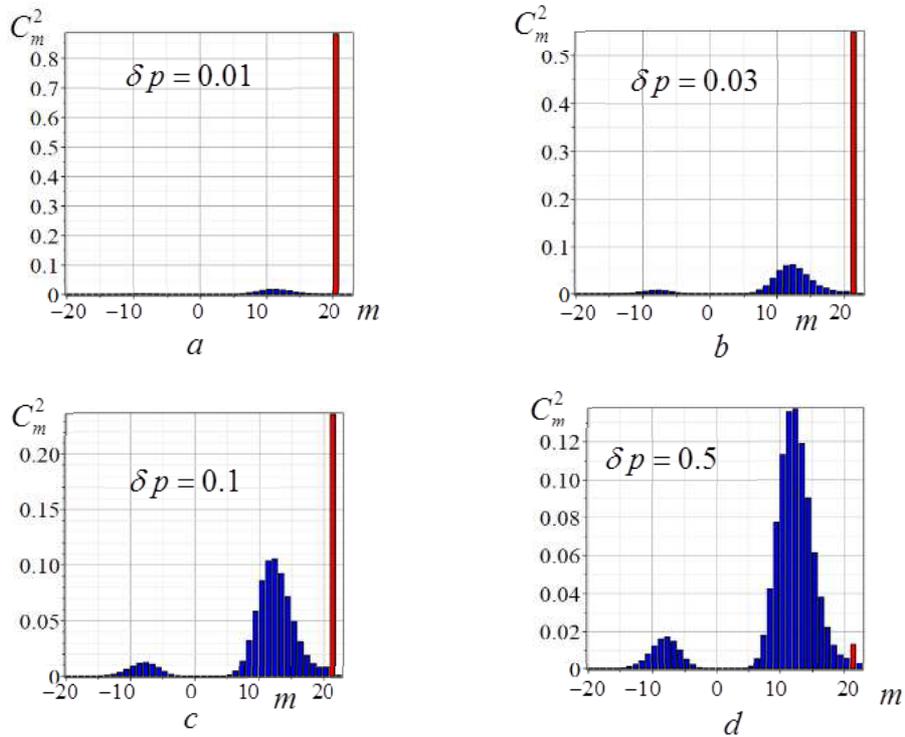

Fig.7 The computer simulation of the intensity redistribution $C_m^2(m)$ over the beam modes $m$ in the state $|20\rangle$ (red color) for different values of the perturbations $\delta p$, $\Omega = 10$

To estimate positions $m_{max}$ of intensity maximum, we assume that the effective perturbation parameter $p$ is very large, e.g. $m/p \ll 1$ and $\Omega < p$ in Eq. (13). Then the position $m_{max}$ of the satellite squared amplitudes $C_m^2$ can be determined by the condition $dC_m^2/d\Omega = 0$. After simple transformations we find

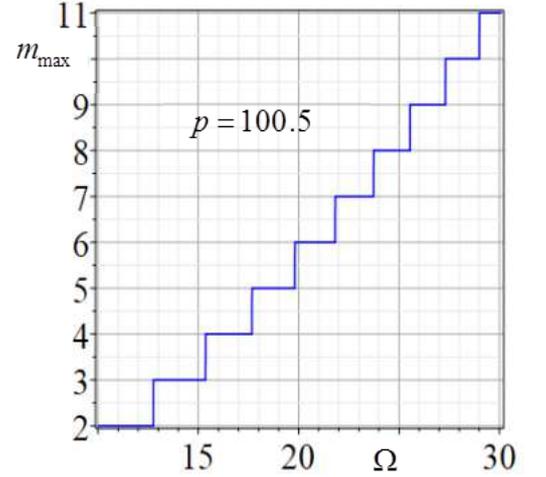

Fig.8 Shift of the of the satellite maximum $m_{max}$ with a change of the scale parameter $\Omega$, $p = 100.5$

$$m_{max} = (round)\frac{p I_1(2\Omega) + 4\Omega^2 I_0(2\Omega)}{4\Omega I_1(2\Omega) + 2p I_0(2\Omega)} \qquad (14)$$

where the operator $(round)$ sets the nearest integer to the fractional number and we used the definition of the modified Bessel function

$$I_m(x) = \sum_{n=0}^{\infty} \frac{(x/2)^{2n+m}}{m!(n+m)!} . \qquad (15)$$

As can be seen in Fig.8, the positions $m_{max}$ of the shifted satellites is nearly directly proportional to the scale parameter $\Omega$. A characteristic feature of the satellites spectral shift is that as the scale parameter increases, the shift of the $C_m^2$ maximum of the satellites approaches asymptotically the topological charge value of the initial vortex beam $m_{max} = M$. This does not mean that OAM has remained unchanged. For example, a small perturbation $\delta p = 0.01$ of the beam state $|100\rangle$ leads to a decrease of the OAM to $\ell_z \approx 16.9$ while a shift of the satellite maximum is $m_{max} = 100$.

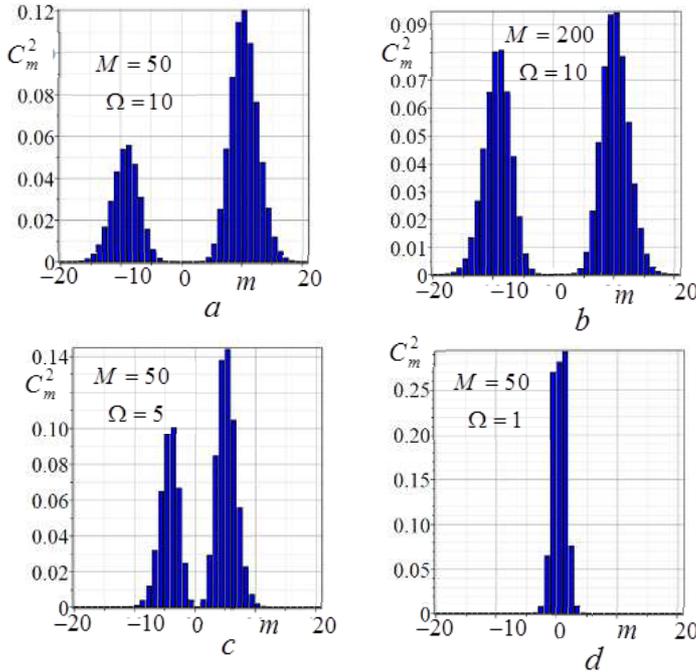

Fig.9 The intensity redistribution $C_m^2(m)$ among the combined beam modes for the perturbation $\delta p = 0.5$ and (a) $M = |50\rangle$, $\Omega = 10$, (b) $M = |200\rangle$, $\Omega = 10$; (c) $M = |50\rangle$, $\Omega = 5$, (d) $M = |50\rangle$, $\Omega = 1$

The computer simulation of the squared amplitudes $C_m^2(m)$ (Eq.14) gives a more accurate pattern of the vortex avalanche instability shown in Fig.9. A small perturbation causes a complete intensity transfer from the initial state to the satellites. The intensity in the side satellites is equalized for very large values of the initial topological charge $M$ (see Fig.9b). As the scale parameter decreases, the side satellites approach each other and when $\Omega \to 1$, they merge (Fig.9a,c,d).

## 4. The experiment and discussion

The measurement of the avalanche instability process of the OAM was accomplished on the basis of a new approach presented in Ref.[26]. The technique was based on the measurement of higher-order intensity moments (see e.g. [17,38,39] and references therein) with the subsequent computer calculation of the amplitudes, phases and OAM of mode beams in the composition of the vortex array. The measurements were accompanied by double independent control of the obtained results. On the one hand, the fractional OAM was measured on the basis of the technique presented in Ref.[25,32] and compared with the main result of measurements of the OAM in the form $\ell_z = \sum_{m=0}^{N-1} m C_m^2$. On the other hand, the initial beam intensity distribution was compared with the intensity of the beam, restored via the measured squared amplitudes. Basic measurements focused on analyzing structure of the intensity distribution $\Im_p(r,\varphi)$.

The characteristic shape of the intensity distribution of the complex amplitude (7) written as

$$\Im_p(r,\varphi) = \Psi_p^* \Psi_p = \sum_{m=0}^{N-1} C_m^2 r^{2m} G^2 + 2 \sum_{\substack{m,j=0,\\m>j}}^{N-1} C_m C_j r^{m+j} \cos\left[(m-j)\varphi\right] G^2 - $$
$$- 2 \sum_{\substack{m,-j=0,\\m>j}}^{N-1} C_j C_m r^{m+j} \sin\left[(m-j)\varphi\right] G^2, \qquad (16)$$

(with the squared amplitudes $C_m^2$ defined by Eq.(13)) makes it possible to measure separately each of the three terms of the equation. Since in our experiment the initial phases of the mode beams in the superposition (7) are not taken into account, it suffices to restrict ourselves to only the first member in Eq.(16).

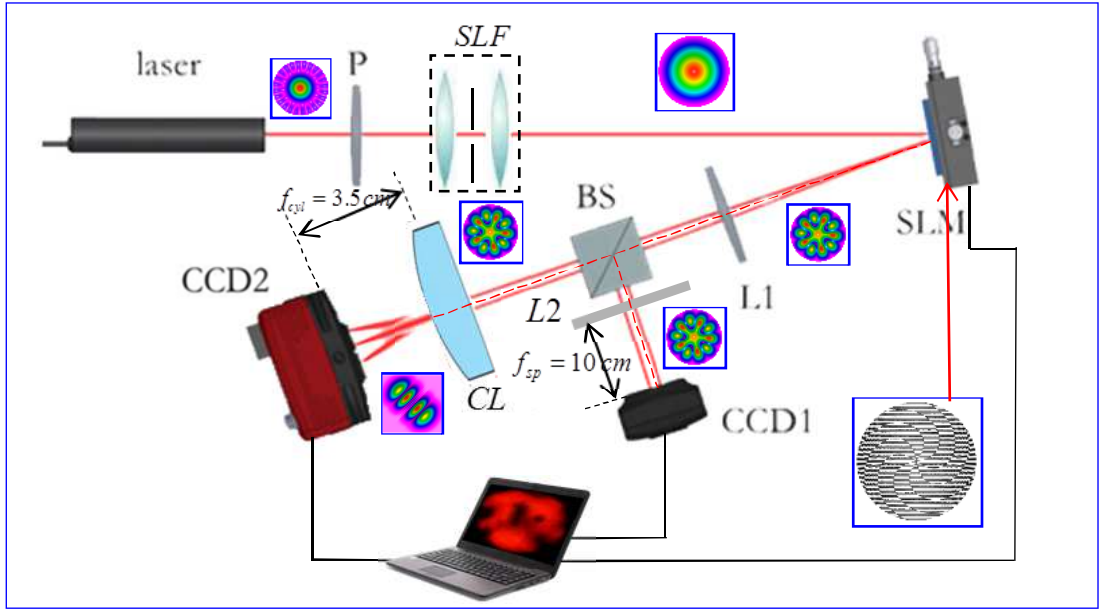

Fig.10 Experimental setup for real-time measuring the vortex and the OAM spectrum, P – polarizer, SLF – space light filter, SLM – space light modulator, L1, L2 – spherical lenses with a focal length $f_{sh}$, BS – beam splitter, CL – cylindrical lent with a focal length $f_{cyl}$, CCD1,2 – CCD camera.

For the experimental analysis, a system of intensity moments $J_{s,q}$ was employed. The intensity moments are written in the standard form [39]

$$J_{s,q} = \iint_S \mu_{s,q}(x,y) \Im(x,y) dx dy / J_{00}, \qquad (17)$$

where $\mu_{s,q}(x,y)$ stands for a function of the intensity moments, $J_{0,0} = \sum_{m=0}^{N-1} C_m^2$ is the beam intensity, $p,q = 0,1,2,3,...$. Since the initial complex amplitude (7) is specified in the axially symmetric basis of Laguerre – Gauss beams, it is convenient to choose the moment function in the form $\mu_s(r) = r^s$, $s = 0,1,2,...N-1$. After substituting Eq. (7) into Eq.(16) we get $N$ liner equations relative to the squared amplitudes $C_m^2$

$$J_s = \sum_{m=0}^{N-1} \frac{(s+m+1)!}{m!} C_m^2, \; p \leq N-1. \qquad (18)$$

The intensity moment is measured experimentally, and the solution of a system of the linear equations gives $N$ squared amplitudes $C_m^2$. Note that the intensity moments $J_p$ are degenerate with respect to the sign of the vortex topological charge; therefore, this version of the technique enables us to measure only combined beams with the same signs of the topological charges.

To measure the OAM and the vortex spectrum (the squared amplitudes and OAM), an experimental setup is shown in Fig.10. The laser beam $\lambda = 0.628\,mcm$ passes through a spatial filter SF and is projected onto a spatial light modulator SLM (*Thorlabs EXULUS-4K1* with the resolution $3840 \times 2160 (4KUHD)$ pixels) that forms an array of $N$ vortex beams with given amplitudes $C_m$ The reflected beam is split into two arms by a beam splitter BS. In the first arm, the beam is projected by a spherical lens onto the plane of the CCD1 camera. The image of the beam cross-section at the beam waist plane is processed by the computer that calculates the intensity moments $J_{p,q}$ in real time. In the course of a computer processing, the intensity moments $J_{p,q}$ are substituted into the equations (18) and the squared amplitudes (the mode intensity) $C_m^2$ of the vortex beams are monitored. For the calculation of the moments $J_{p,q}$, the photocurrent in each pixel of the image is multiplied by the cords $x, y$, the results are summed over the entire image plane, normalized by the beam intensity $J_{00}$ and output to the computer monitor. In the second arm, the beam is focused by a cylindrical lens CL with a focal length $f_{cyl}$ at the plane of the CCD2 camera located at the focal plane of the lens CL. The image is processed by the computer in accordance with the technique of measuring the OAM per photon, described in detail in Ref. [32]. This technique was based on the measurement of the second-order intensity moments $J_{xy}$ at the focal plane of the cylindrical lens CL. The output signals of the SLM modulator and both CCD cameras are synchronized with each other so that the measurement results are displayed on the monitor screen in real time.

At first, we measured the intensity distribution over the mode beams $C_m^2$ and OAM $\ell_z$ of the state $|10\rangle$ with weak perturbations $\delta p = 0.1 \div 0.5$. It was revealed the measured intensity redistributions to be almost the same as that shown the computer simulations in Fig. 7. The avalanche of the state $|10\rangle$ does not occur, the OAM is not lower than $\ell_z \approx 9.48$ at the perturbation $\delta p = 0.5$. According to Fig.1b the state $|10\rangle$ is located in the region of slow changes of the OAM. It is such a smooth change of OAM that is observed by the authors of

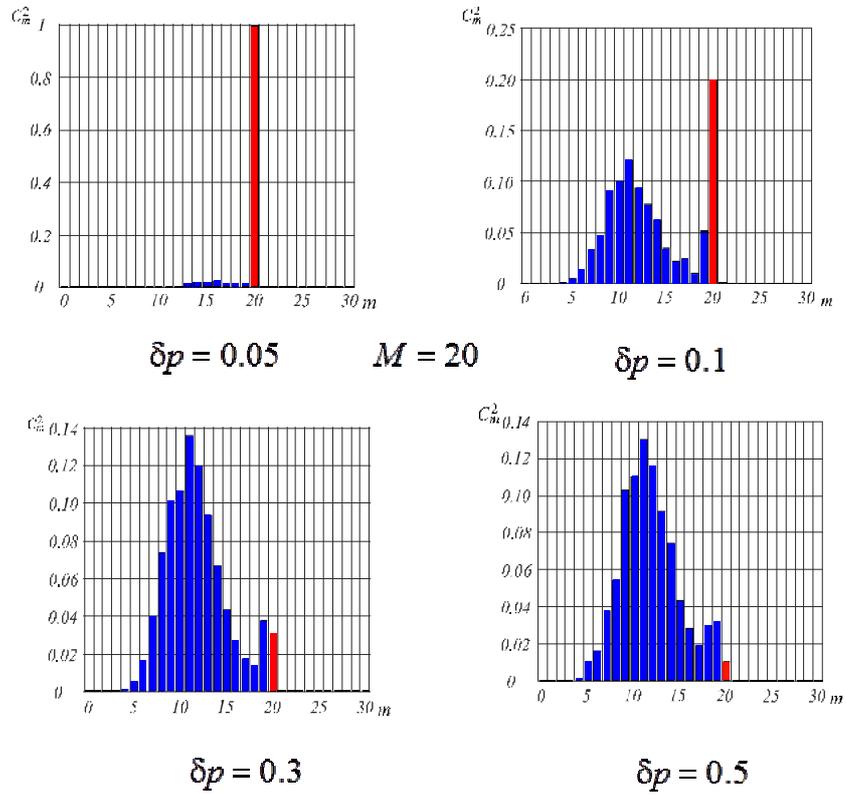

$\delta p = 0.05 \quad M = 20 \quad \delta p = 0.1$

$\delta p = 0.3 \quad \delta p = 0.5$

Fig. 11 The experimental detection of the vortex avalanche in the state $|20\rangle$ caused by a weak perturbation $\delta p$ as the squared vortex amplitudes $C_m^2(m)$ evolution. $\Omega = 10$, the initial state – red.

Ref.[21]. A pronounced avalanche was detected for the state $|20\rangle$ shown in Fig.11. The main problem accompanying the measurement process was that the intensity distribution $C_m^2(m)$ over the mode beams includes both a region of positive and negative topological charges (see Fig.7) while the method used enables us to measure the OAM spectra of the combined beams with the same signs of topological charges. At the same time, the theoretical estimation of the state $|20\rangle$ showed that a contribution of negatively charged vortices to the experimental error will not exceed 7% at the maximal perturbation $\delta p = 0.5$. Note that the experimentally observed breaking-up of the state $|20\rangle$ begins with a perturbation $\delta p = 0.05$ shown in Fig.12a. The maximal avalanche of the state is accompanied by an almost complete intensity transfer into the side satellites (Fig.12d). The intensity distributions $C_m^2(m)$ allow us to plot the OAM spectrum $\ell_z(p)$ shown in Fig.13, curve 1. Far from the integer value $p$, the OAM is approximately $\ell_z \approx 10$. However, in the vicinity of the values

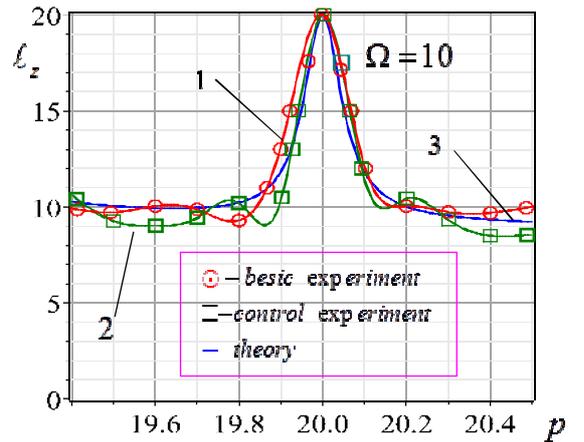

Fig.12 The OAM avalanche instability of the state $|20\rangle$ in the form of the spectral curves $\ell_z(p)$: curve 1 – the basic experiment, curve 2 – the control experiment, curve 3 – the computer simulation of Eq.(9), the scale parameter $\Omega = 10$

$p \in (9.8, 10.2)$, there is a sharp burst of the angular momentum up to $\ell_z = 20$. To exclude any error in the detection of the OAM resonance, we carried out control measurements based on an independent technique developed by the authors of the paper [32]. For this purpose, the second arm of the experimental setup in Fig.11 was used. The intensity momentum $J_{xy}$ of the second order was measured at the focal plane of the cylindrical lens and the value of the OAM was calculated by the formula

$$\ell_z = \frac{4\pi}{\lambda f_{cyl}} J_{xy} = \frac{4\pi}{\lambda f_{cyl}} \iint_S x \cdot y \cdot \Im_{cyl}(x,y) dS / J_{00}, \quad (19)$$

where $\lambda$ is a wavelength, $f_{cyl}$ is a focal length of the cylindrical lens, $\Im_{cyl}(x,y)$ stands for the intensity distribution at the focal plane. The curve 2 in Fig. 12 illustrates the results of the control measurements. A distinct resonance burst with height $h_r \approx 10$ in the vicinity of $p \in (9.8, 10.2)$ is observed both in the basic and control groups of measurements, and they are in a good agreement with the resonant contour of the theoretical curve 3. In order to estimate the total measurement error of the vortex spectra, we compared the intensity distributions of the initial combined beam and the beam restored via the measured squared amplitudes $C_m^2$ employing the correlation degree [23]

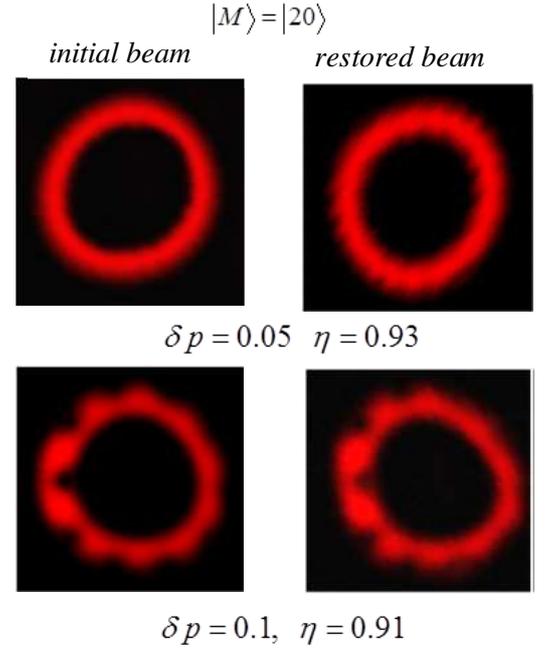

Fig.13 The intensity distributions of the initial and restored combined beams in the state $|20\rangle$ subjected to weak perturbations $\delta p$ and their correlation degree $\eta$.

$$\eta = \iint_S \Im(x,y) \Im_{exp}(x,y) dS / J_{00}^2 \quad (20)$$

where $\Im(x,y)$ and $\Im_{exp}(x,y)$ are the intensity distributions of the initial and restored beams, $J_{00}^{exp}$ stands for the intensity of the initial and restored beams, respectively. Fig. 13 shows the intensity distributions of the original $\Im(x,y)$ and $\Im_{exp}$ restored beams and the degree of correlation between them. The maximum and minimum of the correlation degree covered the interval $\eta \in (0.93, 0.91)$ indicates a high level of reliability of the results obtained.

## 5. Conclusion

Vortices with fractional topological charges are not the exclusive property of optical electromagnetic fields. They can be found in processes of quantum mechanics [45], hydrodynamics [46], acoustics [44], etc. As a rule, such vortices are unstable and rapidly decompose into a superposition of vortices with integer topological charges. In a stable vector state, fractional vortices can exist only in an inhomogeneous medium due to spin-orbit coupling as e.g. in the *A*-phase of superfluid $^3He$ [45] or a space-variant birefringent medium of photonic crystals [18]. But can an optical vortex with an integer topological charge turn into its fractional analogue in free space? In our work, we have shown that such a transformation occurs due to a weak perturbation of a holographic grating responsible for shaping the optical vortex. Moreover, there occurs avalanche collapse of the ground states of the vortex.

We have theoretically and experimentally examined a new model of a combined vortex beam formed by a holographic grating subjected to weak perturbations of its relief. The spectra of squared amplitudes of mode beams and the OAM were analyzed as a function of the perturbation parameter of the grating. We revealed that in the OAM spectrum, there are dips and bursts at the region where the topological charge of the mode beam coincides with the integer part of the perturbation parameter. We have shown that such dips and bursts of the OAM are caused by avalanche decay of the ground state of an optical vortex due to the rapid intensity transfer to spectral satellite modes with positive and negative topological charges. Such a complex vortex composition can be treated in a framework of the model of singular beams bearing fractional-charged vortices. Current experimental measurements were accompanied by the independent control experiments and showed good agreement with theoretical predictions.

The considered effect of the OAM avalanche instability can find application, for example, in monitoring weakly turbulent states and other inhomogeneous media. The fact is that such a medium can be presented as a set of vortices with different scaling. OAM resonances of the combined beam scattered by such a medium change their parameters. Comparison of the OAM spectra of the initial and scattered beams enables us to obtain additional information about the medium state. In addition, the effect of avalanche collapse of mode states should be taken into account when developing holographic gratings for optical systems with compacted information transfer due to modulation of the orbital angular momentum.

**Appendix**

In accordance with the definition [28], the total topological charge is the signed sum of all the vortices threading a large sum including the beam axis, i.e. taking into account Eq.(7) we have

$$S_p = \lim_{R \to \infty} \frac{1}{2\pi} \int_0^{2\pi} d\varphi \frac{\partial}{\partial \varphi} \arg \Psi_p(R) = \lim_{R \to \infty} \frac{\text{Re}}{2\pi} \int_0^{2\pi} d\varphi \frac{\sum_{m=-\infty}^{\infty} \frac{m(\Omega R)^{|m|} e^{im\varphi}}{|m|!(p-m)}}{\sum_{m=-\infty}^{\infty} \frac{(\Omega R)^{|m|} e^{im\varphi}}{|m|!(p-m)}}. \quad (A.1)$$

In the case of a beam containing a finite number of vortices with $m \in (-N, N)$ we have

$$S_p = \lim_{R \to \infty} \frac{\text{Re}}{2\pi} \int_0^{2\pi} d\varphi \frac{\sum_{m=-N}^{N} \frac{m(\Omega R)^{|m|} e^{im\varphi}}{|m|!(p-m)}}{\sum_{m=-N}^{N} \frac{(\Omega R)^{|m|} e^{im\varphi}}{|m|!(p-m)}}. \quad (A.2)$$

There are two possible cases: 1) $p < N$, 2) $p > N$. In the first case $p < N$, a situation is possible when the value of the fractional parameter $p$ approaches the integer value $p \to M < N$. Find the limit of Eq. (A2) when $p \to M$

$$\lim_{p \to N} S_p = \lim_{R \to \infty} \frac{\text{Re}}{2\pi} \lim_{p \to M} \int_0^{2\pi} d\varphi \frac{(p-M) \sum_{\substack{m=-N, \\ m \neq 0}}^{N} \frac{m(\Omega R)^{|m|} e^{im\varphi}}{|m|!(p-m)} + \frac{M(\Omega R)^M e^{iM\varphi}}{M!}}{(p-M) \sum_{m=-N}^{N} \frac{(\Omega R)^{|m|} e^{im\varphi}}{|m|!(p-m)} + \frac{(\Omega R)^M e^{iM\varphi}}{M!}} = M. \quad (A.3)$$

Finally we obtain the total topological charge

$$S_p = \frac{\text{Re}}{2\pi} \int_0^{2\pi} d\varphi \lim_{R\to\infty} \frac{\sum_{m=0}^{N} \frac{-m(\Omega R)^{|m|} e^{-im\varphi}}{|m|!(p+m)} + \sum_{m=1}^{N} \frac{m(\Omega R)^{|m|} e^{im\varphi}}{|m|!(p-m)}}{\sum_{m=0}^{N} \frac{(\Omega R)^{|m|} e^{-im\varphi}}{|m|!(p+m)} + \sum_{m=1}^{N} \frac{(\Omega R)^{|m|} e^{im\varphi}}{|m|!(p-m)}}$$

(A.4)

$$= \frac{\text{Re}}{2\pi} \int_0^{2\pi} d\varphi \frac{N\left(-\frac{e^{-iN\varphi}}{(p+N)} + \frac{e^{iN\varphi}}{(p-N)}\right)}{\left(\frac{e^{-iN\varphi}}{(p+N)} + \frac{e^{iN\varphi}}{(p-N)}\right)} = N,$$

where we use the integral $\int_0^{2\pi} \frac{-a e^{-i2N\varphi} + b}{a e^{-i2N\varphi} + b} d\varphi = 2\pi$.

In the case of a large parameter $p \gg N$ we find

$$\Psi = \frac{\sin p\pi \, e^{ip\pi}}{\pi} \sum_{m=-N}^{N} \frac{(\Omega R)^{|m|} e^{im\varphi}}{|m|!(p-m)} = \frac{\sin p\pi \, e^{ip\pi}}{\pi p} \sum_{m=-N}^{N} \frac{(\Omega R)^{|m|} e^{im\varphi}}{|m|!(1-m/p)} \approx$$

$$\approx \frac{\sin p\pi \, e^{ip\pi}}{\pi p} \sum_{\substack{m=-N,\\m\neq 0}}^{N} \frac{(\Omega R)^{|m|} e^{im\varphi}}{|m|!} (1+m/p) + \frac{\sin p\pi \, e^{ip\pi}}{\pi p} =$$

$$= \frac{\sin p\pi \, e^{ip\pi}}{\pi p} 2\sum_{\substack{m=0,\\m\neq 0}}^{N} \frac{(\Omega R)^m \cos m\varphi}{|m|!} - i2\frac{\sin p\pi \, e^{ip\pi}}{\pi} \sum_{\substack{m=0,\\m\neq 0}}^{N} \frac{m}{p} \frac{(\Omega R)^m \sin m\varphi}{|m|!} + \frac{\sin p\pi \, e^{ip\pi}}{\pi p},$$

so that $S_p = 0$